# HIGHLY EFFICIENT MULTILAYER ORGANIC PURE-BLUE-LIGHT EMITTING DIODES WITH SUBSTITUTED CARBAZOLES COMPOUNDS IN THE EMITTING LAYER.

A. Fischer[1], S. Chénais[1], S. Forget[1], M.-C. Castex[1], D. Adès[2], A. Siove[2], C. Denis[3], P. Maisse[3] and B. Geffroy[3]

[1]Laboratoire de Physique des Lasers (LPL, CNRS), Institut Galilée, Université Paris 13, 93430 Villetaneuse, France.

[2]Biomatériaux et Polymères de Spécialité (BPS/B2OA, CNRS), Institut Galilée, Université Paris 13, Villetaneuse/Faculté de Médecine Lariboisière-St Louis, Université Paris 7, 75010 Paris, France.

[3]Laboratoire Cellules et Composants, CEA/LITEN/DSEN, CEA Saclay, 91191 Gif-sur-Yvette, France.

*e-mail : fischer@galilee.univ-paris13.fr*

**Abstract :**
Bright blue organic light-emitting diodes (OLEDs) based on 1,4,5,8,*N*-pentamethylcarbazole (PMC) and on dimer of *N*-ethylcarbazole (*N,N'*-diethyl-3,3'-bicarbazyl) (DEC) as emitting layers or as dopants in a 4,4'-bis(2,2'-diphenylvinyl)-1,1'-biphenyl (DPVBi) matrix are described. Pure blue-light with the C.I.E. coordinates $x = 0.153$ $y = 0.100$, electroluminescence efficiency $\eta_{EL}$ of 0.4 cd/A, external quantum efficiency $\eta_{ext.}$ of 0.6% and luminance $L$ of 236 cd/m$^2$ (at 60 mA/cm$^2$) were obtained with PMC as an emitter and the 2,9-dimethyl-4,7-diphenyl-1,10-phenantroline (BCP) as a hole-blocking material in five-layer emitting devices. The highest efficiencies $\eta_{EL.}$ of 4.7 cd/A, and $\eta_{ext} = 3.3\%$ were obtained with a four-layer structure and a DPVBi DEC-doped active layer (CIE coordinates $x = 0.158$, $y=0.169$, $\lambda_{peak} = 456$ nm). The $\eta_{ext.}$ value is one the highest reported at this wavelength for blue OLEDs and is related to an internal quantum efficiency up to 20%

## 1. Introduction

Over the last decade, blue Organic Light Emitting Diodes (blue OLEDs) with high efficiencies have attracted considerable attention for their potential applications to the full colour ultra-thin flat panel display [1-7]. Moreover, blue light can be converted into green or red with the use of proper dyes giving the possibility to generate all colours from the blue emitter. This latter property leads to an important simplification in the design of the OLED displays. Very recently, Gebeyehu et al.[4] reported on efficient blue OLEDs with doped transport layers in a p-i-n type structure. From the material point of view, carbazole derivatives have appeared since the nineties to be one of the most promising among various blue-emitting materials [5-9]. For example, efficient blue-light emission has been reported with multilayer diodes using distyrylarylenes (DSA) as emitters: Hosokawa et al.[5] reported an external quantum efficiency ($\eta_{ext}$) of 2.4 % using DSA doped with amino-substitued DSA, whereas Liao et al.[6] recently published a external quantum efficiency of 8.7% thanks to the use of a composite hole transport layer. However, the former exhibited a greenish-blue tint as the peak of the emitted light is 480 nm, and the later is sky-blue with Commission Internationale de l'Eclairage (CIE) coordinates of (x=0.15, y=0.29).

In an earlier work, we have reported that a carbazolic dimer i.e *N,N'*-diethyl-3,3'-bicarbazyl (DEC) is a thermally and electrochemically stable organic compound [10,11] which exhibits interesting pure blue electroluminescence in single double layer and triple layer LED configurations [12-14]. Since the charges injection and the transport were not optimized, quantum efficiencies of the devices remained small ($\eta_{ext.}$ of about 0.1%). Recently, carbazoles such as 4,4'-bis(*N*-carbazolyl)-biphenyl (CBP) [15-18] or dimers and trimers of *N*-octylcarbazole [19] have been used as host materials for triplet emitters in electrophosphorescent devices.

In this letter, we report on blue OLEDs based on the dimer *N,N'*-diethyl-3,3'-bicarbazyl (DEC) and on the novel stable carbazole i.e. 1,4,5,8,*N*-pentamethylcarbazole (PMC). Both can be used either in a five-layer structures with the 2,9-dimethyl-4,7-diphenyl-1,10-phenantroline (bathocuproine, BCP) as a hole-blocking layer or as a doping material in a four-layer structure with a 4,4'-bis(2,2'-diphenylvinyl)-1,1'-biphenyl (DPVBi) matrix. The devices that uses the PMC as an emitting material exhibited the purest bright blue-light with the chromatic coordinates (C.I.E.) x = 0.153, y = 0.100 and with an electroluminescence efficiency (EL) $\eta_{EL}$= 0.4 cd/A and an external quantum efficiency $\eta_{ext}$ of 0.6%. Much higher EL performances were obtained from four-layer structures using DEC as a dopant of DPVBi. For an optimized doping ratio of 2%, devices displayed intense blue-light peaking at 456 nm (CIE coordinates x=0.158 y=0.169), with $\eta_{EL}$ and $\eta_{ext}$ as high as 4.7cd/A and 3.3% (at 10mA/cm$^2$) and with a luminance $L$ of 2835 cd/m2 (at 60mA/cm$^2$). This $\eta_{ext}$ value is among the highest reported for blue OLEDs with comparable CIE coordinates. Moreover, a two-days half life time has been observed for devices tested constinuously with a 9V voltage, under ambient atmosphere and without any encapsulation.

**2.Experimental**

Figure 1 shows the molecular structure of the organic compounds used in this work while figure 2 details the layer structures of the two types of devices (a and b, described later in the text) with the detail of the organic compounds and the thickness of each layer. DEC was synthesized according to the procedure previously described [20]. PMC was recristallized from alcohol prior using. Indium tin oxyde (ITO)-coated glass with sheet resistance of about 15 Ω/□ was obtained from Asahi. Prior using, the glass was cleaned by sonication in a detergent solution and then rinsed in de-ionized water.

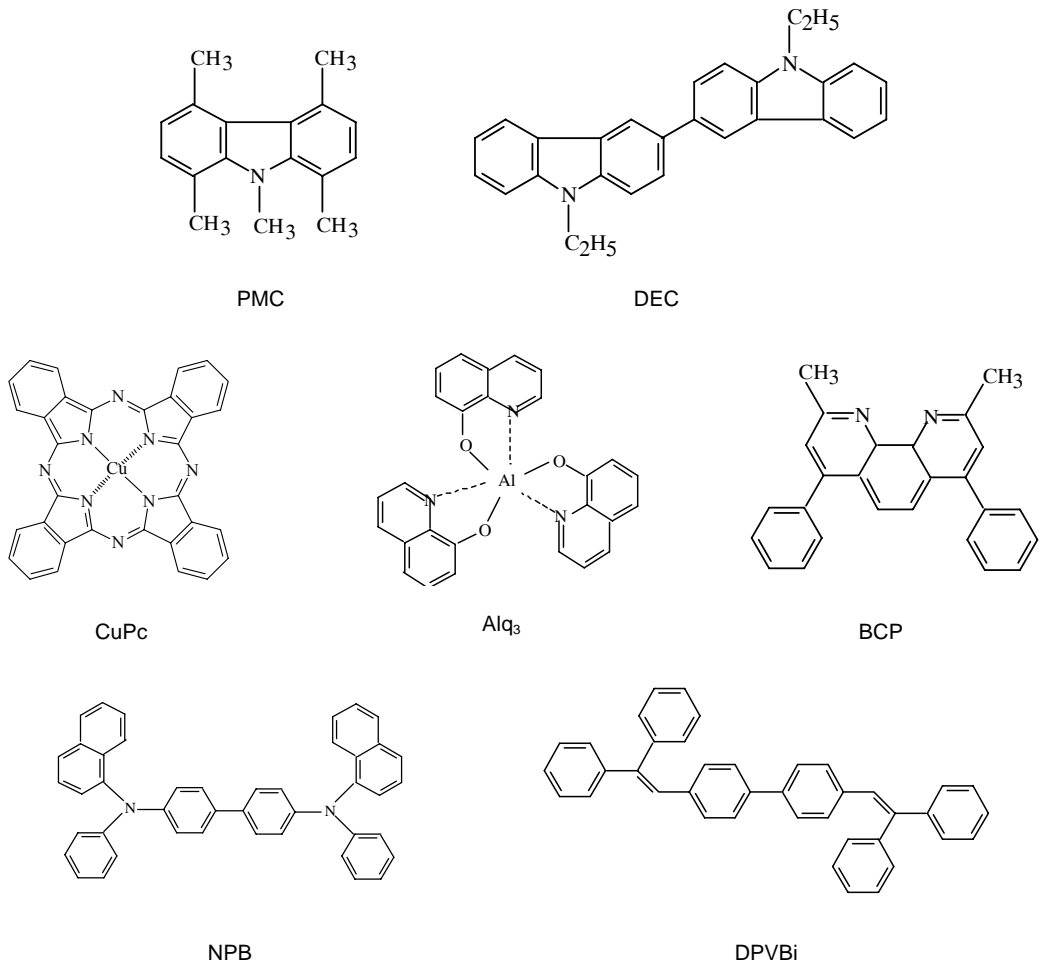

Figure 1 : Molecular Structures of the organic materials used for the Blue-Oled.

| LiF 1.2nm/Al 100nm |
| --- |
| Alq₃ 10nm |
| BCP 10 nm |
| PMC or DEC 50nm |
| NPB 50nm |
| CuPc 10nm |
| ITO Glass |

**(a)**

| LiF 1.2nm/Al 100nm |
| --- |
| Alq₃ 10nm |
| DPVBi (PMC or DEC) 50nm |
| NPB 50nm |
| CuPc 10nm |
| ITO glass |

**(b)**

Figure 2 : OLED structure with the detail of the tickness and the organic material used in each layers; (a) the emissive layer is DEC or PMC, (b) the emissive layer is DPVBi-doped with DEC or with PMC.

The organic compounds are deposited onto the ITO anode by sublimation under high vacuum ($10^{-7}$ Torr) with a rate of 0.2 – 0.3 nm/s. An in-situ quartz crystal was used to monitor the thickness of the vacuum depositions. The active area of each OLED was 0.3 cm$^2$. In (a) configuration (Fig.2(a)), we sequentially deposited onto the ITO anode : CuPc (copper phtalocyanine) as a hole-injecting layer (HIL), (NPB) (*N,N'*-bis(1-naphtyl)-*N,N'*-diphenyl-1,1'-biphenyl-4,4'-diamine) as a hole-transporting layer (HTL), DEC or PMC as an emitting layer (EML), BCP (bathocuproine; 2,9-dimethyl-4,7-diphenyl-1,10-phenantroline,) as a hole-blocking layer (HBL), Alq$_3$ (tris(8-quinolinoato)aluminium) as the electron-transporting layer (ETL), and finally a LiF/Al top cathode. LiF has been found to increase by a 50-fold factor the device efficiency of blue OLEDs DPVBi-based compared to a device with a aluminium cathode only [21]. In another strucure Fig.2 (b), both the DEC (or PMC) layer and BCP layers are replaced by a single DPVBi-DEC (or DPVBi-PMC) doped active layer. The latter is obtained by co-evaporation from two separated evaporation sources. The organic materials (host and dopant) were co-evaporated from two resistively heated evaporation cells which were controlled by a temperature controller and monitored through a thermocouple inserted into the bottom of the cell. By careful control of the evaporation cell temperature, it was possible to precisely deposit a determined amount of dopant molecule dispersed into the emitting layer. The evaporation rates were monitored with two separate quartz monitors and the accuracy of the doping is estimated to 10%.

The doping ratios (in weight %) were optimized to be 2% of DEC (or 5% of PMC) in the DPVBi matrix. EL spectra and chromaticity coordinates were recorded with a PR 650 SpectraScan spectrophotometer. The Current-Voltage and Current-Luminance (I-V and I-L) characteristics of the diodes were measured with a regulated power supply (ACT 10 Fontaine) combined with a multimeter and a 1 cm$^2$ area silicon calibrated photodiode (Hamamatsu). All the measurements were performed under ambient atmosphere.

Figure 3 shows the energy levels for both structures. The HOMO and LUMO levels and the energy band gap $E_g$ for DEC and PMC compounds were deduced from electrochemical measurements of their oxidation and reduction potentials. For DEC, HOMO and LUMO levels were found to be 5.6 eV and 2.5 eV respectively whereas for PMC HOMO and LUMO levels are 5.9 eV and 2.8eV. These values agree quite well with those deduced from the absorption threshold of the optical spectra [22].

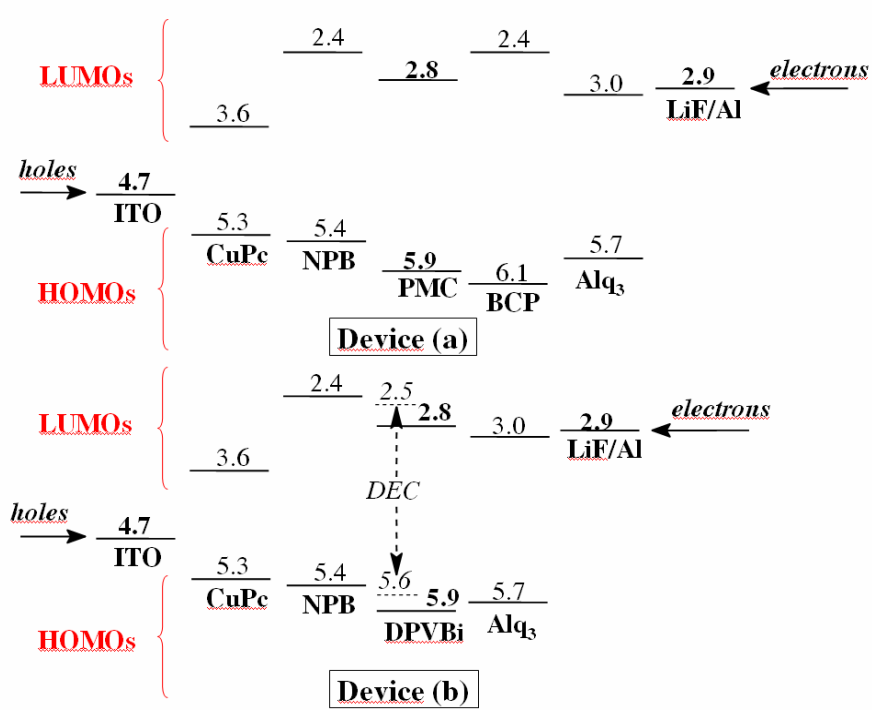

Figure 3 : Energy diagram of the heterostructure with the details of the HOMO and LUMO levels (in eV) (device (a) : for a five-layer structure with the emitting layer made of PMC, and device (b) : for a four-layer configuration in which the emitting layer is a DPVBi matrix doped with DEC.

**3. Results and discussion**

Figure 4 (a) shows the EL spectra of DEC and PMC in the five-layers configuration containing BCP as a hole blocking layer (a-type according to figure 2). The bright blue colour emission of these OLEDs is clearly visible in ambient room light. The DEC EL spectrum is characterised by an emission peak at 424 nm with a 150 nm FWHM (full width at half maximum), while the PMC spectrum is characterized by a narrower peak at 450 nm (FWHM = 80 nm) as expected for this less $\pi$-conjugated structure. Moreover the shoulder at 550 nm in the DEC emission spectrum has been previously observed in monolayer OLEDs [12] and attributed to the creation of excimers or agregates. The EL spectra of carbazole-doped or undoped DPVBi structures (b-type devices according to figure 2) are shown on figure 4 (b). Spectra of the blends show a structureless feature and are almost identical to that of pristine DPVBi. Similar blends have already been used to improve the quantum efficiency of blue-light emitting devices including distyrylarylene doped with amino-substituted derivatives [5] or BCzVBi (an analog of DPVBi in which diphenyl end-groups are replaced by carbazole rings) doped in CBP [7].

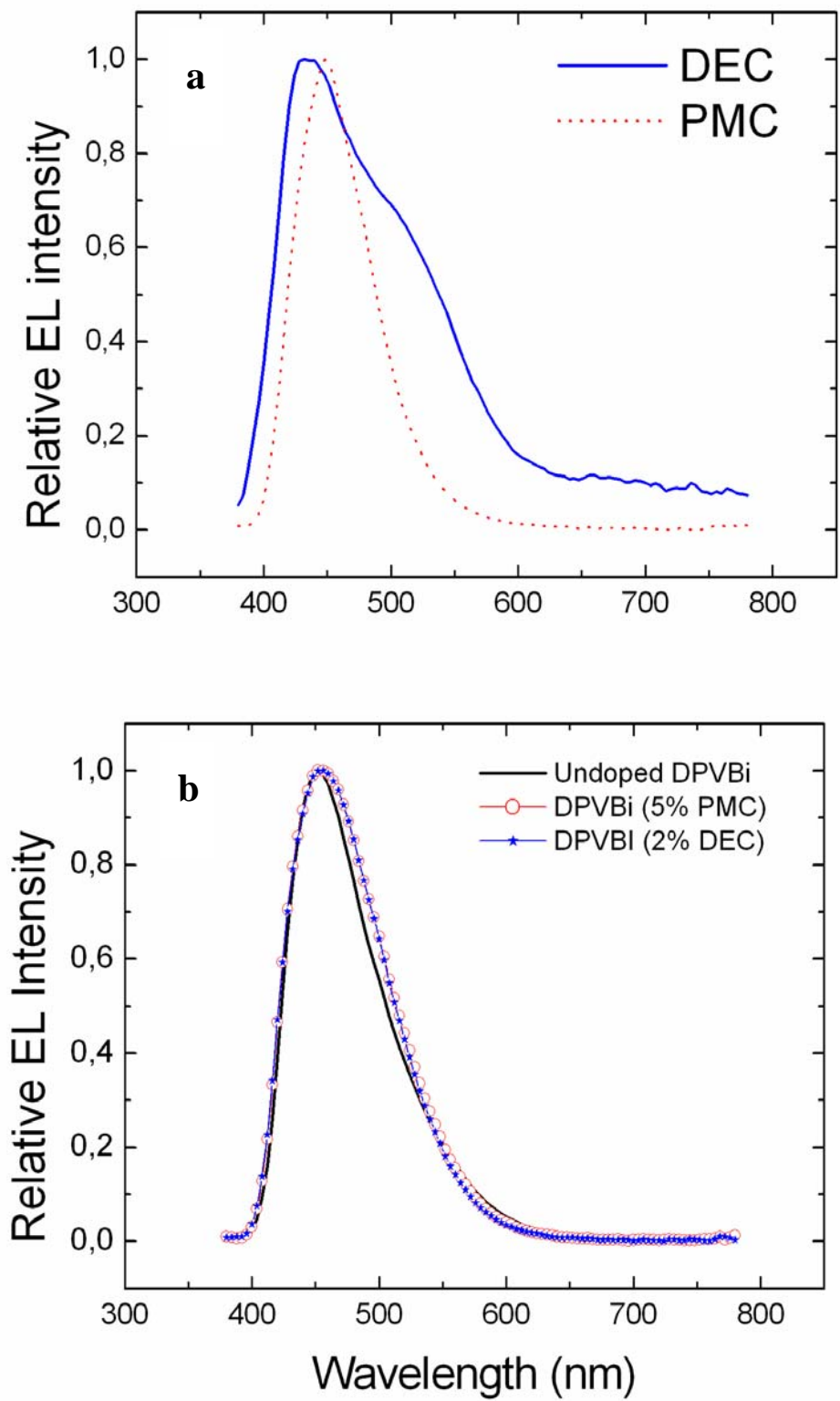

Figure 4 : Electroluminescence spectra of five-layer (**a**) and four-layer (**b**) devices

We plot in the figure 5 a typical exemple of the evolution of the efficiency (in cd/A) with the applied voltage for our devices (here for a b-type device with 5% of PMC in the DPVBi host).

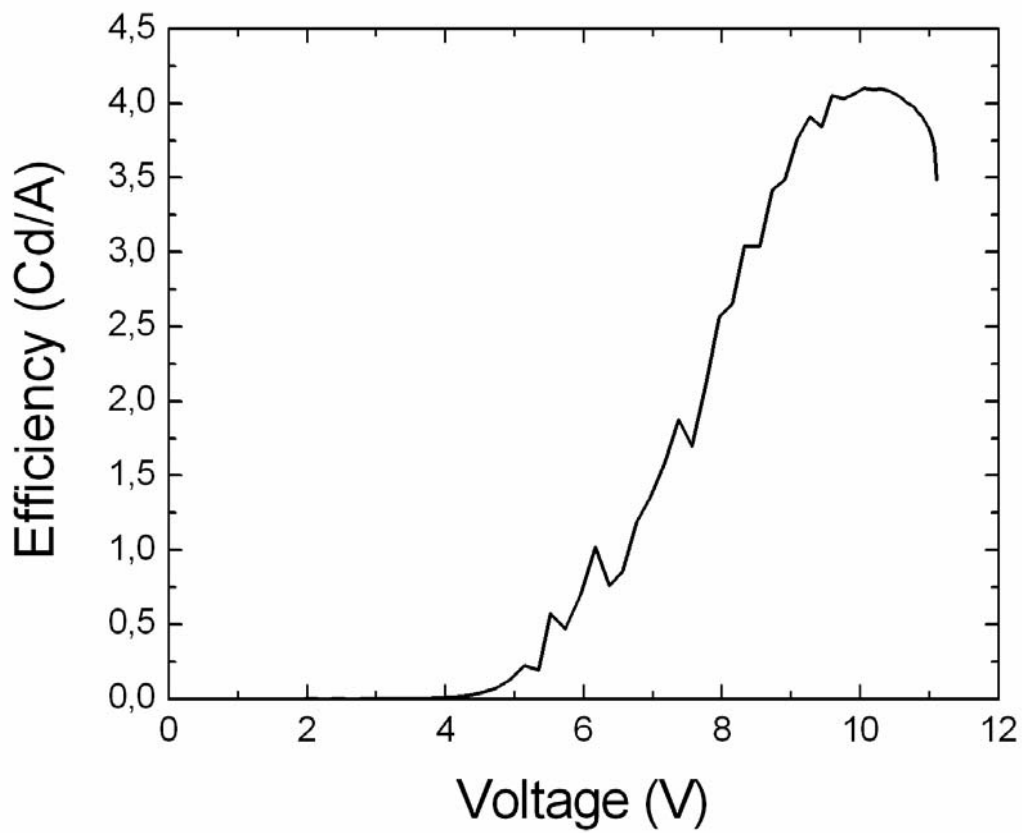

Figure 5 : efficiency (in candela/ampere) versus voltage for a type-b device (5% PMC doped DPVBi)

The electroluminescence parameters for all the devices are presented in the table 1. In optimal DPVBi-carbazole doping ratio (in weight %) the electroluminescence efficiencies of doped devices were improved by a nearly two-fold factor compared to that of nondoped devices. The DPVBi DEC-doped device exhibits a quantum efficiency of $\eta_{ext}$ = 3.3%. This value is to our knowledge one of the highest reported for deep-blue light emitting diodes, close value beeing obtained with a CBP-doped device [7] or with spiro-anthracene or derivative of TPD in pin type device [4].

Table 1 : EL performances of the PMC and the DEC-based multilayer blue-light emitting diodes (at a current density of 10 mA/cm$^2$ ). Column PMC and DEC deal with the structures shown fig.2a while the last 3 columns deal with the fig.2b structures.

|  | Device (a) PMC | Device (a) DEC | Device (b) DPVBi PMC-doped (5%) | Device (b) DPVBi DEC-doped (2%) | Device (b) DPVBi nondoped |
|---|---|---|---|---|---|
| $\eta_{ext}$ (%) | 0.6 | 1.5 | 2.8 | 3.3 | 2.7 |
| $\eta_{EL}$ (cd/A) | 0.4 | n.d [b] | 4.1 | 4.7 | 2.4 |
| $\eta_{power}$ (lm/W) | 0.2 | n.d | 1.2 | 1.3 | 1.2 |
| $L$ (cd/m$^2$)[a] | 236 | n.d | 2279 | 2825 | 1507 |
| C.I.E. $x$ | 0.153 | 0.192 | 0.160 | 0.158 | 0.149 |
| C.I.E. $y$ | 0.100 | 0.209 | 0.176 | 0.169 | 0.112 |

[a] at 60 mA/cm$^2$. [b] non determined

As no special attention to cavity effects was paid in the architecture of the OLED, we assume [23] a classical light output coupling efficiency of $\eta_{out} \approx 1/2n^2 \approx 16$ % for a glass substrate with a n $\approx 1.77$ refractive index of the emissive layer (corresponding to the DEC refractive index extrapolation in the UV range), the internal quantum efficiency $\eta_{int} = \eta_{ext} / \eta_{out}$ of the device is found to be closed to 20%. This means that 20% of the injected electrons give a photon in the material : the other 80% are in a first approach composed of three terms : a) the injected electrons which do not recombine to form an exciton ($\eta_a$), b) the excitons that lead to a triplet state, and consequently are not luminescent ($\eta_b$, around 75% in a rough estimation for non-phosphorescent materials), and c) the exciton, although singlet, that exhibits a non-radiative decay ($\eta_c$). We can deduce from our measurements that the product $\eta_a \cdot \eta_c$ is as high as 82 %. Regarding the LUMO and HOMO levels of the materials involved, an efficient electron-hole balance across the device (b) could results in those very high performances. Furthermore, although the emission of the DVPBi being the major part of the EL spectra and the energy bandgaps of both the host and the guests being equal ($E$g= 3.1 eV), a Förster energy transfer can not be completely ruled out [24], and dopant excitation could arise from a better electron-hole recombination with the carbazole compounds playing the role of the carrier traps according to the trap mechanism [25].

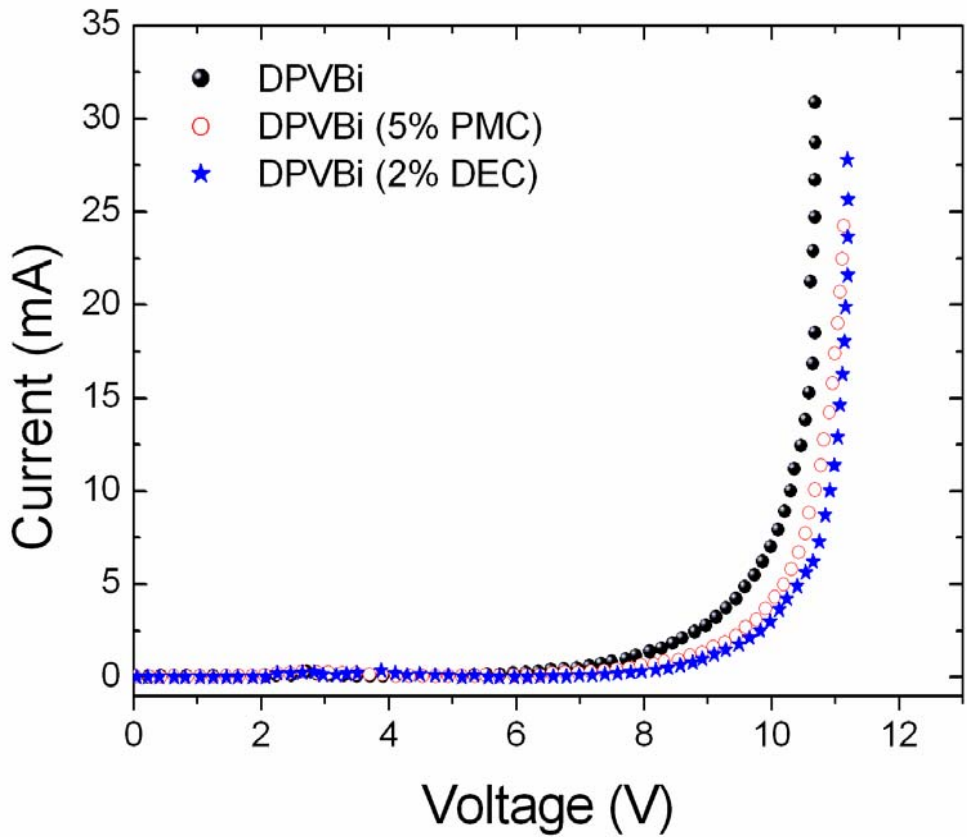

Figure 6 : Current *vs*. Voltage characteristics of the doped and undoped structures.

The electrical characteristics of the 3 host-guest devices (b-type in figure 2), without any doping, 5% PMC-doped and 2% DEC-doped respectively, are shown in figure 6. Every device exhibits a diode behaviour with a threshold value around 8V. Noteworthy, the required voltage for reaching a given current density is higher for the doped devices compared to the nondoped one. This is in agreement with the behaviour of reported devices [24, 25] with dopants working as carrier traps. Indeed, hole mobility is expected to decrease when holes are trapped into dopants, so leading to an increase of the electric resistance of the emitting layer.

Furthermore, it is noteworthy that the external quantum efficiency of the five-layer DEC based-diode (a-type) is one order of magnitude higher than the one obtained for the previously described double-layer devices [12-14], the power conversion efficiency beeing two orders of magnitude higher. This drastic improvement can be easily understood by considering the different HOMO and LUMO levels of the energy diagram shown in figure 3. Progressive steps between ITO and DEC facilitate the holes to diffuse across the junction. The energy barrier between the HOMO level of the DEC and that of the DPVBi or the BCP ($\Delta E \sim 0.3$ to $0.5$ eV) is large enough to block the hole diffusion into the emitting layer and thus to confine the electron-hole recombination in the active layer. Moreover the BCP, DPVBi, DEC and PMC LUMO levels match pretty well and thus favor the electron transport from the cathode. Hence, the radiative recombination is confined inside the emitting layer. Nevertheless, the energy barrier between the LUMO level of the BCP and that of the LiF/Al cathode ($\Delta E=0.6$ eV) needs to be reduced for further improvement of the EL performances. That shows up in the results obtained with DPVBi (table 1). Indeed, this molecule allows a better adjustment of the considered LUMOs ($\Delta E= 0.1$ eV).

The C.I.E. coordinates reported in table 1 show that all the devices emitted a pure blue light and that the DVPBi host tends to take the emission colour of the carbazolic guest. Noteworthy the use of PMC as an emitting layer in a five-layer structure (device shown fig.2(a)) produces a blue light at the frontier of the violet-blue (C.I.E $x = 0.153$, $y = 0.100$). Furthermore, the electroluminescence of the blue OLEDs were tested under ambient atmosphere without any encapsulation, during a continuous two-days test. PMC-based active layer structures appear to live slighty longer than other structures with DEC. During that 2-days test, the active surfaces were progressively polluted with dark spots and EL performances were half lower. Three distinct degradation mechanisms have been identified in small-molecule based OLEDs [26], dark-spot degradation, catastrophic failure and intrinsic degradation. The dark-spot degradation is known to come from the aluminium cathode oxidation or delamination due to moisture. Nevertheless this degradation mode can be reasonably solved by means of adequate control over device fabrication conditions (clean room, glove-box, encapsulation).

In conclusion, highly efficient blue-light-emitting-diodes based on two carbazole molecular derivatives *i.e*. the dimer of *N*-ethylcarbazole (DEC) and the 1,4,5,8,*N*-pentamethyl-carbazole (PMC) as wide-band gap emitters were achieved. We emphasized the advantages of matching both the HOMOs and the LUMOs of the various organic materials in order to improve the electron-hole balance accros the junction. In such a way, efficient multilayer bright-blue light emitting diodes have been fabricated. The DPVBi DEC-doped devices exhibits a state-of-the-art external quantum efficiency of 3.3% related to a high internal efficiency of 20 %, whereas the theoretical maximum of the singlet emission is 25%.